\documentclass{mn2e} 
\usepackage{times}
\input{psfig.sty}

\newif\ifAMStwofonts

\title[The anti-hierarchical growth of supermassive black holes]
      {The anti-hierarchical growth of supermassive black holes}

\author[Merloni] {Andrea Merloni\\Max-Planck-Institut f\"ur Astrophysik,
Karl-Schwarzschild-Strasse 1, D-85741, Garching, Germany }

\date{}

\begin{document}

\maketitle

\label{firstpage}

\begin{abstract}
I present a new method to unveil the history of cosmic accretion and the
build-up of supermassive black holes in the nuclei of galaxies, based
on observations of the evolving radio and (hard) X-ray luminosity
functions of active galactic nuclei. The fundamental plane of black
hole activity discovered by Merloni, Heinz \& Di Matteo (2003), which
defines a universal correlation among black hole mass ($M$), 2-10 keV
X-ray luminosity and 5 GHz radio luminosity is used
as a mass and accretion rate estimator, provided a specific functional
form for the dependency of the X-ray luminosity on the dimensionless
accretion rate $\dot m$ is
assumed. I adopt the local black hole mass
function as derived from the velocity dispersion ($\sigma$) distributions of
nearby galaxies coupled with the $M-\sigma$ relation
as a boundary condition to integrate backwards in time the continuity
equation for the supermassive black holes evolution, neglecting
the role of mergers in shaping up the black hole mass function. 
Under the most general assumption
that, independently on $M$, 
black hole accretion proceeds in a radiatively efficient way
above a certain rate, and in a radiatively inefficient way below, 
the redshift evolution of the black hole mass function
and the black hole accretion rate function (i.e. the distribution of
the Eddington scaled accretion rates for objects of any given mass)
are calculated self-consistently. The only tunable parameters are the overall 
efficiency of extracting gravitational energy from the accreting gas,
$\epsilon$, and the critical ratio of the X-ray to Eddington
luminosity, $L_{\rm 2-10 keV, cr}/L_{\rm Edd}\equiv x_{\rm cr}$, 
at which the transition between
accretion modes takes place. 
For fiducial values of these parameters ($\epsilon=0.1$ and 
$x_{\rm cr}=10^{-3}$), I found that half ($\sim 85$\%) of the local 
black hole mass density was accumulated at redshift $z<1$ ($z<3$),
mostly in radiatively efficient episodes of accretion.
The evolution of the black hole mass
function between $z=0$ and $z\sim 3$ shows clear signs of an 
{\it anti-hierarchical} behaviour: while the majority of the 
most massive objects
($M\ga 10^9$) were already in place at $z\sim 3$, lower mass ones
mainly grew at progressively lower redshift, so that 
the average black hole mass increases with
increasing redshift. Also, the average accretion rate decreases
towards lower redshift. Consequently, 
sources in the radiatively inefficient regime
of accretion only begin to dominate the comoving accretion energy
density in the universe at $z<1$ (with the exact value of $z$ 
depending on $x_{\rm cr}$), while at the peak of the black hole
accretion rate history, radiatively efficient accretion dominates by
almost an order of magnitude. I will discuss the implications of these
results for the efficiency of accretion onto SMBH, the quasars
lifetimes and duty cycles, the history of AGN feedback in the
form of mechanical energy output and, more generally, 
for the cosmological models of structure formation in the universe. 
\end{abstract}

\begin{keywords}
accretion, accretion disks -- black hole
physics -- galaxies: active -- galaxies: evolution -- quasars: general
\end{keywords}

\section{Introduction}
In the last decade, supermassive black holes (hereafter SMBH) in the 
nuclei of galaxies
have been discovered at ever increasing pace, and are now believed to
reside in most of (perhaps all) the bulges of present-day galaxies
(Kormendy \& Richstone 1995; Richstone et al. 1998). Moreover,
there is evidence of a correlation between the mass of the central
black holes and either the mass and luminosity (Kormendy \& Richstone
1995; Magorrian et al. 1998; McLure \& Dunlop 2002; Marconi \& Hunt 2003) or
the velocity dispersion (Ferrarese \&  Merritt 2000; Gebhardt et
al. 2000; Merritt \& Ferrarese 2001; Tremaine et al. 2002) of its host bulges.
This has led to the recognition that the formation and growth of SMBH
and of their host galaxies are related processes, and that
understanding their evolutionary history can provide fundamental
insight into the theories of structure formation in the universe and
into the physical nature of AGN feedback (Silk
\& Rees 1998; Fabian 1999; Kauffmann \& Haehnelt 2000; Umemura 2001;
Granato et al. 2001; Cavaliere \&
Vittorini 2002; Wyithe \& Loeb 2003; Granato et al. 2004).

Within this framework, the most important question that needs to be
answered is when and how the mass currently locked up in 
SMBH was assembled. In practice, this corresponds to asking what is, 
at any given redshift $z$,  the
number of black holes per unit co-moving volume 
per unit (base 10) logarithm of mass, i.e. the  {\it black
  hole mass function} $\phi_{M}(M,z)$ (hereafter BHMF). In the
simplest case of 
purely accretion driven evolution (i.e. assuming mergers do not play
an important role in shaping the BHMF), the
simultaneous knowledge of a {\it black hole accretion rate
distribution function} $\phi_{\dot m}(M,z)$,
 would completely determine the evolutionary
solution, as  the two distributions must be coupled via a
continuity equation (Small \& Blandford 1992; Marconi et al. 2004,
hereafter M04), to be solved given the appropriate boundary conditions.
Moreover, if black holes mainly grew by accretion, a direct link should exist
 between the quasar (QSO) and active galactic nuclei (AGN) 
phenomena, signatures of actively accreting phases,
and nearby SMBH tracing directly the
 entire accumulated mass.
In fact, previous comparisons between direct estimates of the 
local black hole  mass density, $\rho_{\rm BH,0} \equiv 
\rho_{\rm BH}(z=0)\equiv
 \int_0^{\infty} M \phi_{M}(z=0)\,dM$, and the
total energy density liberated by powerful quasars
over the cosmic history, as
originally proposed by Soltan (1982), seem to consistently 
suggest that SMBH mainly grew while they
were active (Salucci et al. 1999; Fabian \& Iwasawa 1999; Yu \&
Tremaine 2002, hereafter YT02; Elvis, Risaliti \& Zamorani 2003; Cowie et
al. 2003; M04). 

From the theoretical point of view, several attempts have been made to
link the evolution of the SMBH population to analytic and
semi-analytic models of structure formation (Efstathiou \& Rees 1988; Haenelt \& Rees 1993;
Haenelt, Natarayan \& Rees 1998; Haiman \& Loeb 1998; Cattaneo,
Haenelt \& Rees 1999; Kauffmann \&
Haenelt 2000; Monaco, Salucci \& Danese 2000; 
Cavaliere \& Vittorini 2000; Volonteri, Haardt \& Madau
2003; Wyithe \& Loeb 2003; Hatziminaoglou et al. 2003; Bromley,
Somerville \& Fabian 2004; Mahamood,
Devriendt \& Silk 2004). 
A somewhat different approach has been followed recently by Di Matteo et
al. (2003), who made use of large cosmological hydrodynamical simulations
to predict the evolutionary history of supermassive black holes growth
and activity. 

 The path from cosmological CDM structure formation models (and/or
 simulations) to
the predicted AGN evolving luminosity functions, however, is dotted with
uncertainties regarding a number of important physical processes.
In general, semi-analytic modelers need to make specific assumptions
(and introduce parameters) to describe gas cooling in dark matter halos,
the fueling of the central nuclear black holes, the physics of the
merger events and the nature of the various feedback
mechanisms (from star formation and/or from AGN).
In order to constrain some of these parameters, all the above models, 
then, need to be tested against observationally
determined signatures of SMBH growth history. The usual test-benches 
are: the evolution of the AGN luminosity functions (in any
specific band), the local black holes mass function, as derived from
the $M-\sigma$ relation, the slope and intercept of the $M-\sigma$
relation itself and the spectrum and intensity of the X-ray background
light, known to be produced by the sum of individual AGN.

Here I would like to propose an alternative approach which is capable
to provide a self-consistent evolutionary picture for supermassive
black holes. Such an approach is largely based on {\it observed} data and
only on a minimal number of theoretical assumptions (and parameters). These
assumptions are only needed to describe the
physics of the innermost accretion process, through which 
black holes shine, and not the more complex interplay between growing 
SMBH and their galactic environment. 
This is an advantage for two reasons: first
of all, the physics of black hole accretion is reasonably well
understood, both theoretically and phenomenologically; second, the
physical properties of the innermost part of an accretion flow, where
the dynamics is almost completely dominated by the strong
gravitational field of the central black holes, should not depend on
cosmology and redshift. 

The approach followed here is similar in spirit to those
of Small \& Blandford (1992); Salucci et al. (1999); 
YT02; M04. More specifically, 
I will make use of the local black hole mass function and
of the evolution of AGN luminosity functions to constrain the
evolutionary history of SMBH. The main novelty of the present work
is the realization that
simultaneous radio and (hard) X-ray observations of accreting black
holes can provide tight constraints on {\it both} the mass and the
accretion rate of an active black hole, through the so-called
``fundamental plane'' relationship for active black holes 
(Merloni, Heinz \& Di Matteo 2003, hereafter MHD03). 
From this,
the history of SMBH growth can be followed in detail up to the
redshift at which reliable X-ray and radio luminosity functions can be
obtained.

The structure of the paper is the following: in
section~\ref{sec:condi} I will describe how the knowledge of the X-ray
and radio luminosity functions of local AGN, and of the local
SMBH mass function can be used to obtain a complete census of the
local SMBH population and activity distribution in the form of an
accretion rate distribution function. A necessary ingredient to perform this
calculation is the functional form that relates the
observed X-ray luminosity of an accreting black hole to its mass and
accretion rate. In section~\ref{sec:modes} I will describe how it is
possible to use our theoretical and phenomenological knowledge of the
different modes of accretion onto a black hole to obtain such a
relation. In section~\ref{sec:red}, the method to calculate the
redshift evolution of both black hole mass and accretion  rate
functions is described, and its basic assumptions clearly spelled
out. Section~\ref{sec:res} is then devoted to the analysis of the most
important results of the calculations and of the main properties of
the evolving SMBH population between $z=0$ and $z\sim
3$. A more general discussion of the implication of these results is
presented in section~\ref{sec:disc}. Finally, I draw my conclusions in 
section~\ref{sec:conc}. 

Throughout this paper, we adopt a background cosmological model
in accordance with the Wilkinson Microwave Anisotropy Probe ({\it
  WMAP}) experiment. The model has zero spatial
curvature, a cosmological constant, $\Omega_{\Lambda} =0.71$ a
cosmological constant $H_0 = 70$ km $s^{-1}$, dominated by
cold dark matter with $\Omega_{m} = 0.29$ and $\Omega_b =0.047$
(Spergel et al. 2003).

\section{The importance of Estimating the conditional Radio/X-ray AGN luminosity
  function}
\label{sec:condi}
Large optical surveys carried out in recent years (Boyle et al. 2000;
Fan et al. 2001; Wolf et al. 2003) probe the evolution
of the QSO luminosity function up to high redshift, and all
agree in establishing a strong
rise in their activity from the local universe up to redshift $z\sim
2$
and a decline above $z\sim 3$. However, inferring the properties of
the entire class of accreting supermassive black holes from
samples selected in a single waveband can be misleading, in particular
in
the case of the optical band, where obscuration effects can be
significant.
Indeed, Barger et al. (2003) have clearly shown that optically
selected
broad-line AGN and QSO form only about one third of the X-ray
background: hard X-ray selected samples, therefore, provide a
more direct probe of SMBH activity (see also Cattaneo \& Bernardi
2003).  Recent works by Hosokawa (2004) and
M04 have also demonstrated that the redshift
evolution of the hard X-ray luminosity function better describes the
growth history of accreting supermassive black holes. Discrepancies
with the results obtained from optically selected QSO luminosity
functions for the average accretion efficiency and the local
black hole mass density can be understood by taking into account the
luminosity dependence of both obscuration and
 bolometric corrections in the different
bands (see e.g. Ueda et al. 2003; Cattaneo \& Bernardi 2003;
Hosokawa 2004; M04).

Similarly, hard X-ray emission reveals fundamental properties of an
accreting black hole:
In a recent paper (MHD03) it has been
shown that  if we define the instantaneous state of
activity of a black hole of mass $M$ (in units of solar masses), by the
radio (at 5 GHz, $L_{\rm R}$) and hard (in the 2-10 keV band, $L_{\rm
  X}$) X-ray luminosity of its
compact core, and represent such an object as a point in the
three-dimensional space ($\log L_{\rm R},\log L_{\rm X},\log M$), all 
black holes (either of stellar mass or supermassive)
will lie preferentially on a plane (the ``fundamental plane'' of
black hole activity), described by the following
equation:
\begin{equation}
\log L_{\rm R}=(0.60^{+0.11}_{-0.11}) \log L_{\rm X}
+(0.78^{+0.11}_{-0.09}) \log M + 7.33^{+4.05}_{-4.07}.
\label{eq:fp}
\end{equation}
Equation (\ref{eq:fp}) can be inverted to  relate BH masses 
to observed nuclear radio and X-ray luminosities:
\begin{eqnarray}
\label{eq:masses}
 \log M & \simeq &  1.28 (\log L_{\rm R} - 0.60 \log L_{\rm
  X}) -9.34 \pm 1.06 \\ \nonumber 
&\equiv& g(\log L_{\rm R}, \log L_{\rm X}).
\end{eqnarray}
This is an entirely empirical relation, and as such is independent on
any accretion (or jet) physical model. It shows, however, that 
disc and jet emission from 
active black holes of any mass, from galactic X-ray binary
sources to the most powerful quasars, are physically and
observationally correlated phenomena. Moreover, as the fundamental
plane relationship is obeyed by the intrinsic hard X-ray luminosities, it is
basically unaffected by absorption, and therefore largely independent
on the validity of any specific unification scheme for AGN. 

One of the consequences of this relationship is that, in an ideal case, 
the {\it conditional radio/X-ray} luminosity
function of active black holes, i.e. the number of sources per unit
co-moving volume 
per unit logarithm of radio and X-ray luminosity, 
$\Psi_{\rm C}(L_{\rm R},L_{\rm X})$,
 could be used to reconstruct the mass function of the underlying 
black hole population. The importance of such a possibility should
not be underestimated. 
For example, it could provide an alternative way to study the
demography of the SMBH population (at any redshift) to be compared
with what obtained from the $M-\sigma$ relation (which the Sloan
Digital Sky Survey will largely contribute to, see e.g. McLure \&
Dunlop 2004), or with any other analysis based on 
different mass estimators (see e.g. Vestergaard 2004). In alternative,
such comparisons can be used to test the redshift evolution of the
fundamental plane relationship itself.

Although the future goal for solving the problem at hand
should therefore be identified with the
study of large multi-wavelength (X-ray and radio in particular) SMBH samples,
nevertheless I will
argue here that it is still possible to make some progress with the
currently available pieces of information.
In fact, the lack of the exact knowledge of the conditional radio/X-ray
AGN luminosity function can be (at least partially) 
superseded, given the two separate radio,
$\phi_{\rm R}(L_{\rm R},z)$,  and X-ray, $\phi_{\rm X}(L_{\rm X},z)$,
luminosity functions at redshift $z$, and an independent estimate of the 
black hole mass function, $\phi_{\rm M}(M,z)$ at the same redshift. 
By taking into account the fundamental plane relationship, we have
that the conditional luminosity function $\Psi_{\rm C}$ has to 
satisfy the following integral constraints:

\begin{equation}
\label{eq:int_x} 
\phi_{\rm X}(L_{\rm X}) d\log L_{\rm X}=\int_{L_{\rm R,min}}^{\infty} \Psi_{\rm C}(L_{\rm X}, L_{\rm R}) d\log L_{\rm R}
\end{equation}

\begin{equation}
\label{eq:int_r}
\phi_{\rm R}(L_{\rm R}) d\log L_{\rm R}=\int_{L_{\rm X,min}}^{\infty} \Psi_{\rm C}(L_{\rm X}, L_{\rm R}) d\log L_{\rm X}
\end{equation}

\begin{equation}
\label{eq:int_m}
\phi_{\rm M}(M) d\log M= \int\!\!\!\int_{\log M<g<\log M+d\log M} 
\!\!\!\!\!\!\!\!\!\!\!\!\!\!\!\!\!\!\!\!\!\!\!\!\!\!\!\!\!\!\!\!\!\!\!
\!\!\!\!\!\!\!\!\!\!\!\!\!
\Psi_{\rm C} (L_{\rm X}, L_{\rm R}) d\log L_{\rm R} d\log L_{\rm X},
\end{equation}
where $g(L_{\rm R},L_{\rm X})$ is defined in equation (\ref{eq:masses}).
In the above formulae, the lower end of the X-ray and radio luminosity
functions, $L_{\rm X,min}$ and $L_{\rm R,min}$, should be chosen in such a
way as to give the same total number of objects with mass larger than a
certain minimal value $M_{\rm min}$, as computed by integrating the BHMF.

In the following, I will make the assumption that any function $\Psi_{\rm
  C} (L_{\rm X}, L_{\rm R})$ that satisfies equations
  (\ref{eq:int_x}), (\ref{eq:int_r}) and (\ref{eq:int_m}) can be
  regarded as the true conditional radio/X-ray luminosity function. To
  begin with, I will show in the next section how to use the above
  formalism to derive informations about the state of activity of the
  local SMBH population, in the specific form of its accretion rate
  distribution function.

\subsection{The census of the local population}
\label{sec:census}
Using the integral constraints provided by
eqs. (\ref{eq:int_x}-\ref{eq:int_m}), it is
possible to deduce a conditional radio/X-ray AGN luminosity function
at redshift zero. In order to do that, a specific choice of the
observed luminosity (and mass) functions needs to be done.
I will adopt the following:
\begin{itemize}
\item{For the hard X-ray luminosity function (HXLF) I use the recently
    estimated one of Ueda et al. (2003). This is arguably the most
    complete luminosity function in the 2-10 keV spectral band,
    spanning the luminosity 
    range of $10^{41.5} - 10^{46.5}$ erg s$^{-1}$, corrected for
    absorption.}
\item{The 5GHz radio luminosity function (RLF) of AGN 
is here derived from the lower frequency one of Willott et al. (2001),
obtained from the 3CRR, 6CE and 7CRS complete samples, assuming for
the radio spectral index a constant value $\alpha_{\rm R}=0.7$ to
rescale the luminosities. Although other
determinations of the local AGN RLF at higher frequencies are
available (see e.g. Sadler et al. 2002), the Willott et al. one is
probably the most accurate to date up to relatively high redshifts,
and this will be instrumental in studying the growth history of the
SMBH population (see section~\ref{sec:red}).}
\item{The local black hole mass function is instead estimated following Aller
    and Richstone (2002). They derived it
 from the local luminosity functions of galaxies of
    different morphologies given in Marzke et al. (1994), together
    with the empirical relationships among total luminosity, bulge
    luminosity and velocity dispersion. However, differently from Aller \&
    Richstone (2002), we do take into account the effects of the 
scatter in the $M-\sigma$ relation\footnote{I will assume throughout
  the paper that the black hole mass--velocity dispersion relation is
  given by $\log M_{\rm BH}=8.18+4.02 \log
(\sigma/200{\rm km}\,{\rm s}^{-1})$, as discussed in Tremaine et al. (2002).}, in the same way as described in 
Yu \& Lu (2004). Although the
    overall effect of considering such a scatter affects the value of
    the total black
    hole mass density only marginally (a dispersion of 0.27 dex
    results in an increase of the local black hole mass density of a
    factor of 1.2, see YT02), nevertheless, the shape
    of the local black hole mass function is strongly affected, as 
demonstrated by M04.}
\end{itemize}

Starting with an initial guess for $\Psi_{\rm C}$, we proceed via
successive iterations, minimizing the
differences between the projections of the conditional luminosity 
function onto the X-ray and radio luminosity axes and the observed
luminosity functions, until we obtain a conditional LF that simultaneously
satisfies the integral constraints given by eq.~(\ref{eq:int_x}),
(\ref{eq:int_r}) and (\ref{eq:int_m}).

Once such an estimate of the conditional luminosity function is found, 
it is possible to derive the local distribution of 
the second fundamental physical parameter that
characterizes any active black hole: its accretion
rate in units of Eddington luminosity, 
\begin{equation}
\label{eq:eff}
\dot m \equiv \epsilon \dot M c^2 / L_{\rm Edd}
\end{equation} 
($\epsilon$ is the accretion efficiency, see section~\ref{sec:eff}). 
Such an inversion, however, is model
dependent, as it depends on the choice of which spectral energy
distribution should correspond to any specific couple of fundamental
parameters $M$ and $\dot m$. In practice, it corresponds to the choice
of the accretion mode of a SMBH of given mass and 2-10 keV
luminosity. This can be done by choosing a specific functional
form $L_{\rm X}=L_{\rm X}(M,\dot m)$,
as I will describe in more detail in the next 
section. 

For the moment, suffice it to say that 
the main results concerning the local SMBH population are summarized
by the black solid lines in Figures~\ref{fig:mf} and \ref{fig:mdotf},
showing the $z=0.1$ black hole mass and accretion rate functions.
The population of local black holes is dominated by sources shining,
in the X-ray band, below $10^{-3}$ of the Eddington luminosity
(assuming a 10\% efficiency, see below). This is in agreement with the
average value found using the X-ray luminosity function of Seyfert 1
galaxies by Page (2001). A more detailed view of the local accreting
SMBH population can be obtained by studying the mean dimensionless accretion rate
as a function of black hole mass, defined as:

\begin{equation}
\label{eq:mdotave}
\langle \dot m (M) \rangle  = \frac{\int_{M}^{\infty} \dot m (M)
\phi_M(M) dM}{\int_{M}^{\infty} \phi_M(M) dM}.
\end{equation}

\begin{figure}
\psfig{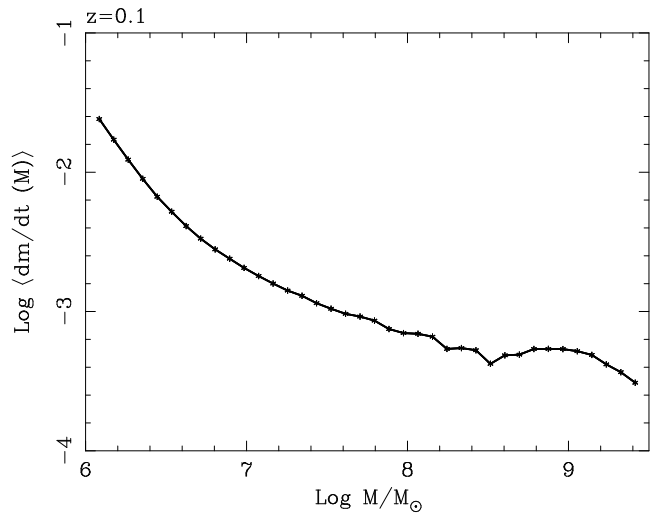}
\caption{The mean accretion rate (in units of Eddington) $\langle \dot
  m (M) \rangle$ as a function of black hole mass for local ($z=0.1$) accreting
  black holes.}
\label{fig:mdot_m}
\end{figure}

\begin{figure}
\psfig{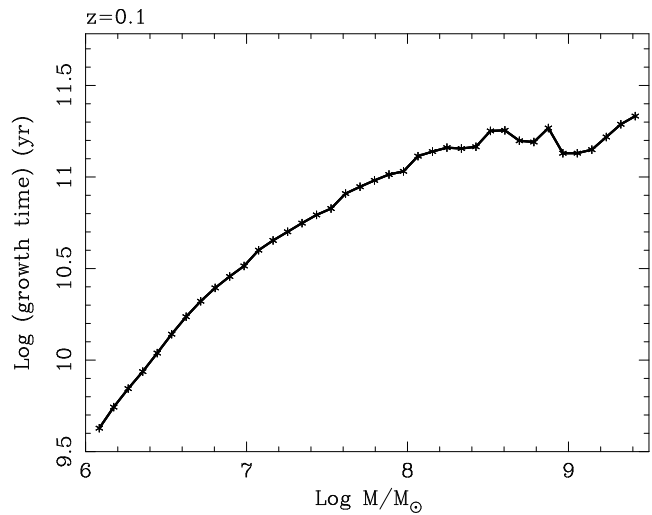}
\caption{The {\it instantaneous growth
rate} of local supermassive black holes, 
defined as $M/\langle \dot M (M) \rangle$
is plotted as a function of black hole mass $M$.}
\label{fig:growth_time}
\end{figure}

In figure~\ref{fig:mdot_m} such a quantity is plotted versus the SMBH
mass at $z=0.1$. It is clear that the mean accretion rate is a strong
function of black hole mass, with small black holes accreting at a
higher rate. Similarly, if we define the {\it instantaneous growth
rate} of a SMBH of mass $M$ as $M/\langle \dot M (M) \rangle$, we see
in figure~\ref{fig:growth_time} that the growth time is very large
(about one order of magnitude larger than the Hubble time) for
the more massive holes, a result confirmed by the {\it SDSS} optical
study of 23,000 local AGN (Heckman et al. 2004). 
This implies that the very high mass
SMBH must have formed at significantly higher redshift, as we will see
in detail in  section~\ref{sec:res}, where the redshift evolution of
the SMBH population
will be calculated.

Before entering into a detailed description
 of the results on the redshift evolution of
the SMBH population obtained using the local distributions 
as a boundary condition, I will discuss  in the next section
the main theoretical assumptions underlying such a calculation.

\section{Theoretical assumptions}
\label{sec:modes}
\subsection{Accretion efficiency}
\label{sec:eff}
The accretion efficiency $\epsilon$, appearing in
equation~(\ref{eq:eff}) represents the 
efficiency with which gravitational energy of the matter infalling
onto the black hole can be extracted, regardless of it being
transformed into radiation or not. It is therefore an upper limit to
the radiative efficiency, $\epsilon_{\rm rad}$, and a function of the
inner boundary condition of the accretion flow only. The closest the
innermost stable circular orbit (ISCO)
of the accreting gas is to the event horizon,
the higher the accretion efficiency. In the classical general
relativistic case of test particles, the position of the ISCO 
depends on the dimensionless angular momentum of the
black holes, and the corresponding efficiency varies between 
$\epsilon\simeq 0.06$ for non-rotating holes and $\epsilon \simeq
0.42$ for maximally rotating Kerr black holes \cite{nt73}. Recently,
both YT02, comparing the local black hole mass function with the
black hole mass density accreted during luminous QSO phases, 
and Elvis, Risaliti \& Zamorani (2002), comparing the X-ray
background intensity with the local SMBH mass density, came to the
conclusion that most supermassive black holes must rapidly
spinning, i.e. they must have accreted at an average radiative 
efficiency higher
than the canonical value of 10\%, so that $\epsilon \ga \epsilon_{\rm
rad} \ga 0.1$. 
Such a conclusion, however, have been questioned by M04 
on the basis of a revised estimate of the local SMBH
mass function and of the hard X-ray luminosity function of AGN.

In what follows, I will assume that the accretion efficiency is a
constant, regardless of the nature of the specific accretion mode of
each SMBH. The {\it radiative} efficiency, instead, will be a function
of the accretion rate, expressed though the dependence of the
observable 2-10 keV luminosity on $\dot m$ and of the bolometric
correction, as described below.

\subsection{Accretion mode transition and bolometric corrections}
The relevance of the different (theoretical) accretion modes for the
various AGN populations is still a matter of open debate. Here I will
follow the approach of MHD03 and try to maximize the
amount of information on the issue by
comparing active black holes of different masses.

Galactic (stellar mass) black holes in X-ray binaries, 
either of transient or persistent
nature, are commonly observed undergoing so-called transitions,
i.e. dramatic changes in their spectral and variability properties
(I will adopt here the terminology of McClintock \& Remillard 2004,
which the reader is referred to for a recent review).
There are at least three well defined  spectral states. In the {\it
  low/hard} state the emission is dominated by a hard X-ray power-law
 with an exponential cutoff at about few 100 keV. The spectrum of the
 {\it thermal dominant (or high/soft) } state, 
instead, is dominated by a thermal component 
likely originated in a standard Shakura \& Sunyaev (1973) accretion disc,
while in the {\it steep power-law (or very high)} 
state, usually associated with a source's
highest flux level, both a thermal and a steep power-law 
component substantially contribute to the spectrum. Maccarone (2003)
has shown that the hard-to-soft transition in these systems generally
occurs at X-ray luminosities (in the 2-10 keV band) of about $x_{\rm
  cr} \simeq 5\times 10^{-3}$. This transition is also accompanied
by a ``quenching'' of the steady radio emission observed in the
low/hard state. 

Given the many similarities between the high-energy spectra of
galactic black holes and AGN, a similar phenomenology has long been
searched for in active supermassive black holes.
The reader is referred to MHD03 for a thorough
discussion on the scale-invariant properties of black holes coupled 
accretion/jet system, and on 
how to constrain theoretical accretion 
models for the different classes of objects
on the basis of the observed fundamental plane 
correlation coefficients. There we  showed that, for black
holes of any mass 
characterized by $x \equiv L_{\rm X}/L_{\rm Edd} \la 0.01$, the 
fundamental plane relation is consistent with the most general
theoretical relation between radio emission, mass and accretion
rate expected from synchrotron emitting jets (regardless of their
detailed geometrical and kinematical properties), provided that the
X-ray emitting flow is radiatively inefficient (RIAF; for a recent 
review about RIAF, see Narayan 2002 or Quataert 2003). In this case, $L_{\rm
  X} \propto \dot m^{2.3}$, and the radio
luminosity satisfies: 
$L_{\rm R} \propto \dot{m}^{1.38} M^{1.38} = \dot{M}^{1.38}$,
i.e., $L_{\rm R}$ scales with the {\em physical accretion rate} only.

On the other hand, the most luminous sources, as those falling into the
standard definition of Quasars and broad lined AGN,
 must be accreting at an higher
rate, close to the Eddington one. Also their spectral energy
distribution, usually dominated by the so-called Big Blue Bump
(BBB: quasi thermal UV emission most likely from an optically thick standard accretion
disc, see e.g. Malkan 1983; Laor 1990), 
indicates that above a certain critical X-ray to Eddington ratio
$x_{\rm cr}$, an accretion mode transition should take place  to what
is usually described as a standard, geometrically thin and optically
thick accretion disc (Shakura \& Sunyaev 1973). More recently, 
Maccarone, Gallo \& Fender (2003), analysing 
the same AGN sample of MHD03, have found evidence of a
connection between the radio-quiet AGN and X-ray binaries in the
thermal dominant state. Also the transition luminosity, $x_{\rm cr}$, has
been found to be consistent with the hard-to-soft transition of
galactic black holes.  

The scaling of the the hard X-ray luminosity
with the accretion rate in this high $\dot m$ regime, 
$L_{\rm X} \propto \dot m^q$, 
is not straightforwardly predicted by the standard accretion disc 
theory, as the origin of the hard X-ray emission itself is not
self-consistently predicted by the theory (but see, for example,
Merloni 2003). 
On the other hand, observational studies of the 
spectral energy distributions of a large 
number of QSO and AGN and comparisons of 
X-ray and optical luminosity functions of AGN 
can be used to put constraints 
on the value of $q$ \cite{ued03}.

In light of these facts, I will adopt here the simplest possible
functional form 
for the $L_{\rm X}/L_{\rm Edd}$ vs. $\dot m$ function of a broken
power-law, bridging the low accretion rate (radiatively inefficient)
regime and the high
accretion rate one. For radiatively efficient sources, we should
assume, by definition, that the bolometric luminosity is simply
proportional to the accretion rate, $L_{\rm bol}\propto \dot m$. Then,
using the fitting formula for the X-ray to bolometric luminosity
ratio of M04, I obtain $x\propto \dot m^{0.76}$. Summarizing, the
universal accretion mode function adopted here has the following scalings:  
\begin{equation}
\label{eq:xmdot}
x \propto \left\{ 
\begin{array}{ll}
\dot m^{2.3} & x \la x_{\rm cr} \nonumber \\
\dot m^{0.76} & x > x_{\rm cr}  \\
\end{array} \right.
\end{equation}
The overall
normalization is found by imposing continuity at $x_{\rm cr}$ and 
that in the radiatively efficient
regime $\log x/ (0.76 \log \dot m)  = -1.5$, such that AGN at the peak
of the mass distribution have X-ray to optical ratios consistent with
observations (see e.g. Vignali, Brandt \& Schneider 2003).

\section{The redshift evolution of the black hole mass and accretion
  rate functions}
\label{sec:red}

In section~\ref{sec:census} I have briefly described the general method by
which to derive the local supermassive black holes mass and accretion rate
distribution functions given the luminosity functions of AGN in both
radio and X-ray bands (down to a sufficiently low luminosity in each
band in order to match the integrated number densities).
From these, the redshift evolution of the SMBH population can be
computed integrating {\it backwards} the continuity equation that
describes SMBH evolution driven by accretion only  (Small \& Blandford
1992; Steed \& Weinberg 2004;
Hosokawa 2004):
\begin{equation}
\label{eq:continuity}
\frac{\partial \phi_{\rm M}(M,t)}{\partial t}+\frac{\partial
[\phi_{\rm M}(M,t) \cdot \langle\dot M(M,t)\rangle]}{\partial M}=0,
\end{equation}
where the mean accretion rate as a function of black hole mass and
time, $\langle \dot M\rangle$ can be calculated directly from the
accretion rate distribution function at time $t$. By
setting the right hand side of equation~(\ref{eq:continuity}) to zero,
I have implicitly assumed that mergers are unimportant for the black holes
growth history. A thorough examination of the possible consequences
for SMBH growth of the inclusion of additional merger or direct
formation terms in equation~(\ref{eq:continuity}) can be found in
YT02, Hosokawa (2004) or Menou \& Haiman (2004). 

In practice, starting from the BHMF and the accretion rate function
 at a given redshift $z$, 
it is possible to derive the new black hole mass function at redshift
 $z+dz$, 
$\phi_{\rm M}(M, z+dz)$, by just subtracting the mass accreted in the
 time interval $dt=dz(dt/dz)$ calculated according to the accretion
 rate function of redshift $z$. This new BHMF can then be
used together with the radio and X-ray luminosity functions at the
 same redshift,
$\phi_{\rm R}(L_{\rm R}, z+dz)$ and $\phi_{\rm X}(L_{\rm X},
z+dz)$, to obtain the new conditional luminosity function 
$\Psi_{\rm C}(L_{\rm R},L_{\rm X}, z+dz)$, 
and therefore the new accretion rate function,
and so on. Thus, the local BHMF, as determined 
independently from the galaxy velocity
dispersion distribution and the $M-\sigma$ relation, and the local
accretion rate distribution function,
as derived from the conditional radio/X-ray LF and a specific
accretion modes scenario, i.e. the function $x(\dot m)$, 
can be used together as a boundary condition to
integrate eq.~(\ref{eq:continuity}) up to the redshift where the HXLF
 and the RLF of AGN can be reliably estimated. At each redshift, I
 discard all SMBH whose mass has decreased below 
$M_{\rm min}=10^6 M_{\odot}$, so
 that nothing can be said, within such a scheme, 
about the formation of the SMBH
 seeds. Furthermore, at every redshift, the values of $L_{\rm X,min}$
 and $L_{\rm R,min}$ (see equations \ref{eq:int_x}, \ref{eq:int_r})
 are increased to take into account the loss of SMBH below our
 threshold. In this way, at every redshift 
the number of objects above a certain minimal
 mass is always equal to the number of radio and X-ray sources above
 the corresponding limiting luminosities.

The final result will depend on the following assumptions:
\begin{itemize}
\item{The X-ray and the Radio luminosity functions of AGN, if
    extrapolated down to low enough luminosity, do indeed
describe the same class of objects (SMBH), and this class of objects is
assumed to be described by the mass function $\phi_{\rm M}$. Apart
from physical arguments invoking the same inner engine for radio and X-ray
selected AGN, there are indeed strong similarities between the cosmic
evolution of radio sources and of X-ray and optically selected AGN
(Dunlop 1998; Willott et al. 2001), a further hint of the correctness
of this assumption.}
\item{The fundamental plane of black hole activity is the same at all
redshifts, and the correlation coefficients are the same as those 
observed at redshift zero. As remarked in the introduction, this is a
natural consequence of the fact that the fundamental plane is a
relationship among physical quantities in the innermost region of the
coupled accretion flow-jet system. There, the strong black hole
gravitational field dominates, and cosmological evolution is
negligible. 
However, it should be noted here that the radio
emission from steep spectrum AGN, whose RLF we adopt here, 
may be sensitive to properties of the interstellar and intergalactic
medium which should indeed depend on redshift. Ideally, the redshift
evolution of the AGN radio cores luminosity function should be used,
as the fundamental plane (\ref{eq:fp}) is a relationship between 
core luminosities only.}
\item{The evolution with redshift of both $\phi_{\rm R}$ and $\phi_{\rm X}$
is known. This is certainly more accurate for the hard X-ray
luminosity function, and in the following we adopt the luminosity
dependent density evolution (LDDE) model of the 2-10 keV luminosity function
 derived by Ueda et al. (2003). The analytic approximation to the HXLF
 is described in the Appendix.

On the other hand, the high redshift evolution of the RLF is much 
less certain. The best estimate to date is probably that by Willott et
al. (2001). This is based on three redshift surveys of
flux-limited samples of steep spectrum sources selected at low
frequencies. By selecting only steep spectrum sources the authors made
sure that the effect of strongly beamed sources (which have typically
flat spectrum) were minimized, especially for the highest redshift
sources. However, the fundamental plane relationship was determined by
including both flat and steep spectrum sources (but excluding beamed
objects, see MHD3), and the selection was
made at higher frequencies. Despite this potential source of uncertainty, 
I will here adopt the Willott et
al. (2001) parameterization, assuming a uniform radio spectral index
$\alpha_{\rm R}=0.7$ 
to rescale the fluxes. Previous studies of redshift evolution of
the RLF of flat and steep spectrum sources (Dunlop \& Peacock 1990) 
have shown that although steep spectrum sources dominate number counts, the
two populations display similar redshift evolution. The analytical
approximation to the RLF of Willott et al. (2001) used here can be
found in the Appendix.}

\item{The function $L_{\rm X}/L_{\rm Edd}=x(\dot m)$ does not depend on
the black hole mass and is expressible as broken power-law: at low
accretion rates black holes accrete in a radiatively inefficient way,
and $x\propto \dot m^{2.3}$, while at high accretion rates BH are
radiatively efficient, with the bulk of the emission being radiated in
the optical/UV bands, so that $x\propto \dot m^{0.76}$ (see section~\ref{sec:modes}).}
\end{itemize}

The whole history of supermassive black hole growth can then be
reconstructed from the evolution of X-ray and radio AGN luminosity
functions with just three free parameters: the accretion efficiency
$\epsilon$, the value of the critical ratio $x_{\rm cr}$ that
separates the radiatively inefficient and efficient regimes, and the
corresponding critical accretion rate or, equivalently, an X-ray 
bolometric correction.

\begin{figure}
\psfig{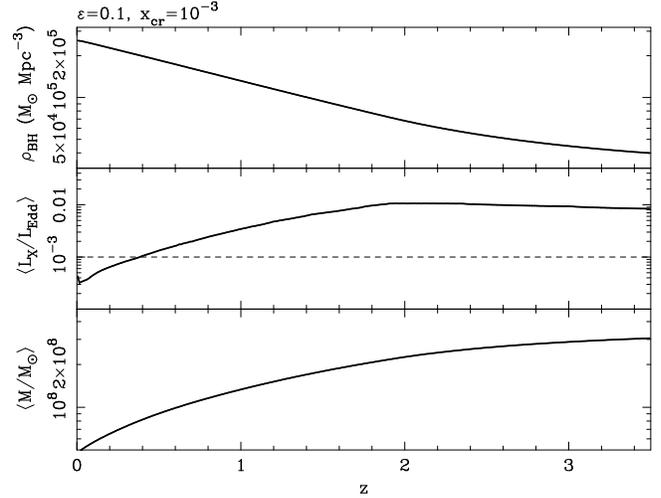}
\caption{Redshift evolution of the comoving black hole mass density
  $\rho_{\rm BH}$ (top panel), of the average X-ray to Eddington
  ratio, $\langle L_{\rm X}/L_{\rm Edd}\rangle$ (middle panel), and of
  the average SMBH mass, in units of solar masses (bottom panel). The
  dashed horizontal line on the middle panel marks the adopted value
  of the critical accretion rate, $x_{\rm cr}$, separating 
radiatively inefficient accretion from radiatively efficient one: most
  of the SMBH growth, therefore, took place during episodes of radiatively
  efficient accretion (see also Fig.~\ref{fig:zevmdot}).}
\label{fig:zevmass}
\end{figure}

\begin{figure}
\psfig{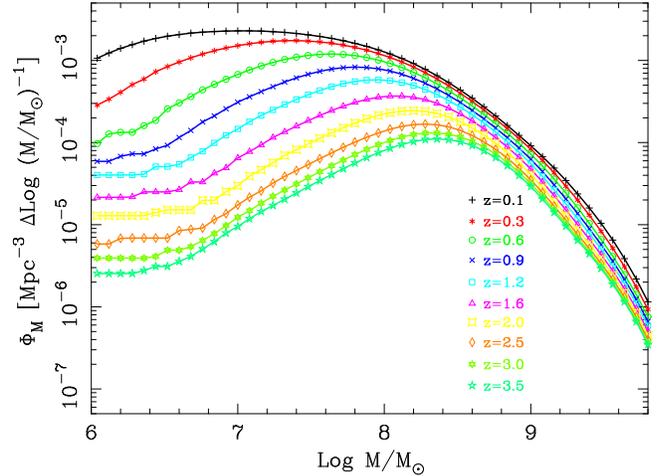}
\caption{Redshift evolution of the SMBH mass function (BHMF), from redshift
  3.5 till redshift 0.1. Different colors and symbols correspond to
    different redshift bins.}
\label{fig:mf}
\end{figure}

\section{Results}
\label{sec:res}
The family of all possible outcomes of the calculation outlined above
 is obtained by varying parameters in the
relatively narrow range of physically and phenomenologically realistic
values: $0.06<\epsilon<0.42$ and $10^{-4} \la x_{\rm cr} \la
 10^{-2}$. 
However, the uncertainties on the observed luminosity functions, the
high-redshift radio one in particular, do not allow to put tight
constraints on these parameters. More interesting, instead, is
the overall trend emerging from this calculation
for the cosmic evolution of the SMBH
population and of its activity level. For this reason, here 
I discuss in detail the results of a calculation performed assuming
$\epsilon=0.1$ and $x_{\rm cr}=10^{-3}$, leaving a discussion of the
consequences of a different choice of parameters 
to section \ref{sec:explo}.

Figure~\ref{fig:zevmass} shows, in the upper panel, 
the redshift evolution of the comoving
black hole mass density from $z=3.5$ to $z=0$. 
The local black hole mass density is $\rho_{\rm BH}(z=0)=2.6 \times
10^5\, M_{\odot}$ Mpc$^{-3}$, a value consistent with that obtained by
YT02, as already discussed by Aller and Richstone,
(2002), while the SMBH mass density at redshift 3  is about six times
lower: $\rho_{\rm BH}(z=3)=4.5 \times
10^4\, M_{\odot}$ Mpc$^{-3}$.
Half of the total black hole mass density was accumulated at
redshift $z<1$. This is consistent with the most recent results from
X-ray Background (XRB) studies.
After the deep {\it Chandra} and {\it XMM} surveys have
revealed the redshift distribution of the obscured sources that most
contribute to the XRB (Alexander et al. 2001; Barger et
al. 2002; Mainieri et al. 2002; Rosati et al. 2002; Hasinger 2003; 
Fiore et al. 2003), the newest synthesis models 
seem to suggest that indeed a substantial fraction of the locally
measured mass density 
of SMBH (maybe up to 50\%)  was accumulated in (obscured)\footnote{It
  is worth emphasizing that both the fundamental plane
  relationship and the HXLF used here rely on {\it absorption
    corrected} 2-10 keV luminosities, and are therefore unaffected by
  Compton thin absorption. Therefore throughout the paper no
  distinction is made (and is possible) between obscured and
  unobscured sources.} low-mass
AGN (with $M\la 10^8 M_{\odot}$) at $z<1$ (see e.g. Gandhi \& Fabian
2003, or Fabian 2003 and references therein). This is also consistent
with the general picture for the BHMF evolution discussed below.

The middle panel of Figure~\ref{fig:zevmass}
 shows the evolution of the average X-ray to Eddington rate,
defined as:
\begin{equation}
\langle \frac{L_{\rm x}}{L_{\rm Edd}}(z)\rangle \equiv \langle x(z)
\rangle  \equiv
\frac{\int_{x_{\rm min}}^{\infty} x \phi_{x}(z)\, dx}
{\int_{x_{\rm min}}^{\infty} \phi_{x}(z)\, dx}.
\end{equation}
Here $\phi_{x}(z)$ is the X-ray to Eddington ratio function,
representing the number of sources per unit co-moving volume per unit
logarithm of the X-ray to Eddington ratio. It is simply related to the
accretion rate function thorough the monotonic function  of 
equation~(\ref{eq:xmdot}): $\phi_{x}(z)=\phi_{\dot m}(z) d\dot m/dx$.
In the same panel, the dashed horizontal line marks the value of
$x_{\rm cr}$. The average accretion rate increases by almost an order
of magnitude from redshift zero until $z \sim 2$, where the luminosity
density of the hard X-ray selected sources peaks, and then levels off.
Supermassive black holes were more active in the past, in the sense
that their average dimensionless accretion rate was higher at higher
redshift. A similar conclusion was drawn by Small \& Blandford (1992),
who adopted a phenomenological approach close to the one followed here,
and by Haiman \& Menou (2002) and Menci et al. (2003) in
the framework of semi-analytic models for structure formation in cold
dark matter universes (see discussion below, \S~\ref{sec:disc}).

\begin{figure}
\psfig{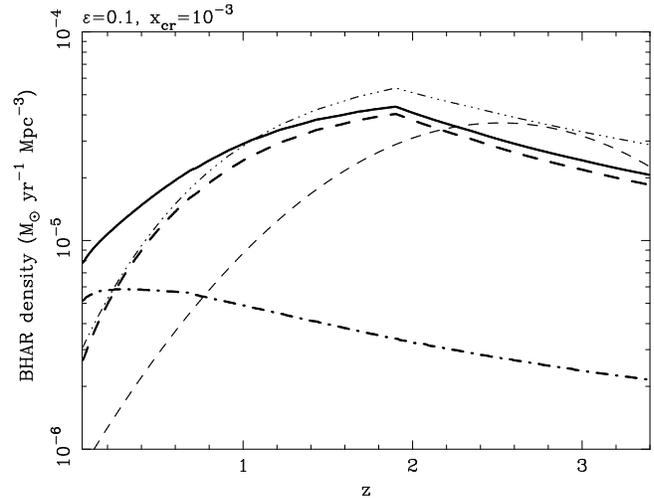}
\caption{Comparison of the evolution of comoving black hole accretion
  rate densities computed according to different prescriptions. The
  thick solid line shows the results from this work, calculated for
  $\epsilon=0.1$ and $x_{\rm cr}=10^{-3}$. The thin triple dotted-dashed line,
  instead, shows the results derived from the hard X-ray luminosity function alone, assuming radiative
efficiency for all sources and a fixed bolometric correction (see
  e.g. M04). Thin
  dashed line shows the evolution of the comoving accretion rate
  density calculated from
the evolution of the QSO luminosity function of Boyle et al. (2000)
with the  bolometric correction of Elvis et al. (1994). Thick dashed and dot-dashed lines in the same plot represent the
accretion rate density for sources above and below the critical rate,
respectively.}
\label{fig:zevmdot}
\end{figure}

The bottom panel of Fig.~\ref{fig:zevmass},
shows the average black hole mass, 
\begin{equation}
\langle M(z) \rangle  \equiv 
\frac{\int_{M_{\rm min}}^{\infty} M \phi_{M}(z)\, dM}{\int_{M_{\rm min}}^{\infty} \phi_{M}(z)\, dM},
\end{equation}
where at every redshift we have taken $M_{\rm min}=10^6 M_{\odot}$.
The {\it anti-hierarchical} (Granato et al. 2001, 2004; M04)
nature of supermassive black holes growth is summarized by this plot,
showing the increase of the average black hole mass with increasing
redshift.

More specifically, the evolution of the {\it shape} of the black hole
mass function, which in turn determines the evolution of $\langle M \rangle$,
is shown in Figure~\ref{fig:mf}. As opposed to the standard picture of
hierarchical mass build up of dark matter halos in CDM cosmologies,
supermassive black holes growing by accretion between $z\sim 3$ and
now have a mass function
which is more and more dominated by largest mass objects at higher and
higher redshift, at least up to the limit where we can trust
the evolution of our parametrized X-ray and radio luminosity
functions. 
As it is indeed emerging from the study
of the QSO population of the SDSS (Vestergaard 2004; McLure \& Dunlop
2004), most of the more massive black
holes ($M \ga 10^9$) were already in place at $z\sim 3$.

The history of accretion activity is instead summarized in 
figure~\ref{fig:zevmdot}, where I plot the comoving black hole
accretion rate (BHAR) density, (thick solid line). The
thick dashed and dot-dashed lines in the same plot represent the
accretion rate density for sources above and below the critical rate,
respectively. For comparison, also plotted are the evolution of the
comoving accretion rate density calculated by M04 
from the hard X-ray luminosity function alone, assuming high radiative
efficiency for all sources (thin
triple dotted-dashed line); 
and the corresponding quantity, calculated instead from
the evolution of the QSO luminosity function of Boyle et al. (2000)
with the  bolometric correction of Elvis et al. (1994) (thin dashed line).
The accretion rate history is therefore dominated by sources with high
radiative efficiency, which explains why many authors were indeed able
to explain most of the local SMBH population as remnants of bright AGN
phases (see e.g. YT02). However, due to the progressive
decrease of the average accretion rate with time, as shown in
Fig.~\ref{fig:zevmass}, the population of 
accreting supermassive black holes at $z\la
0.5$ is dominated by radiatively inefficient sources. For this reason,
the evolution of the BHAR density from $z=0$ to $z=1$ is less rapid
than what would be inferred assuming that all growing SMBH are high
$\dot m$ objects. The consequences of this fact for the history of AGN
feedback are discussed in section~\ref{sec:feedb}.

\begin{figure}
\psfig{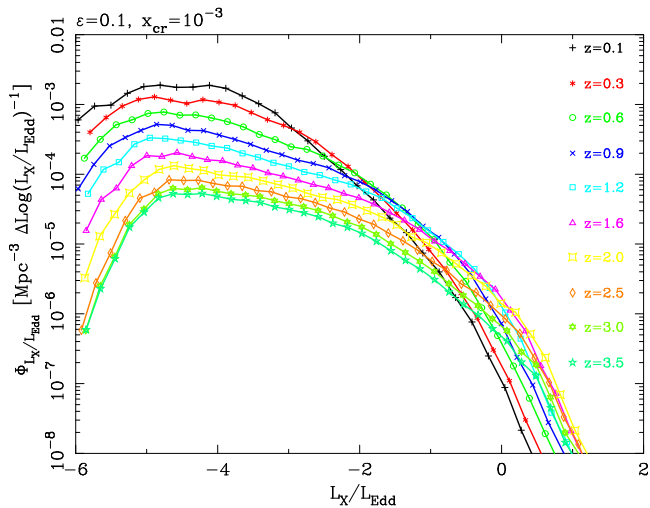}
\caption{Redshift evolution of the SMBH accretion rate function,  
from redshift  3.5 till redshift 0.1. 
Different colors and symbols correspond to different redshift bins.}
\label{fig:mdotf}
\end{figure}

Finally, in Figure~\ref{fig:mdotf}, the evolution of the black hole
accretion rate (expressed here as X-ray to Eddington ratio, $L_{\rm
  X}/L_{\rm Edd}$) function is plotted for different redshift intervals,
from $z=0.1$ to $z=3.5$. While the number of sources accreting at low
rates increases monotonically with decreasing redshift, the situation
is different for rapidly accreting objects. High accretion rate
sources (which should correspond to QSO and bright AGNs) rapidly
increase in number with increasing redshift. The cut-off redshift,
above which the number of sources declines again, is a function of the
typical X-ray to Eddington ratio, being lower for lower accretion rate 
sources.

The combined evolution of mass and accretion rate
functions derived here is the cause of the strong trend
observed in deep X-ray selected samples (Cowie et al. 2003; Hasinger
2003; Fiore et al. 2003), where progressively 
lower luminosity AGN reach their maximal space density
at progressively lower redshifts.

\subsection{Exploring the parameter space: constraints on accretion efficiency}
\label{sec:explo}
As discussed already, recent works by YT02, Elvis et
al. (2002), M04, have shown that 
a direct comparison between local SMBH mass density and
the AGN/QSO energy density integrated over cosmic time can be used to
constrain accretion physics and, in particular the mean radiative
efficiency of all accreting SMBH throughout their history. In doing so, 
the obvious inequality should be satisfied, that
the local $\rho_{\rm BH,0}$ be always larger than (or at most equal to)
the total mass density accreted onto active black holes. If this is
not the case, the simplest solution to the problem is to assume
 a higher accretion (and  radiative) efficiency, as indeed argued by
 Elvis et al. (2002) and YT02.

\begin{figure}
\psfig{figure=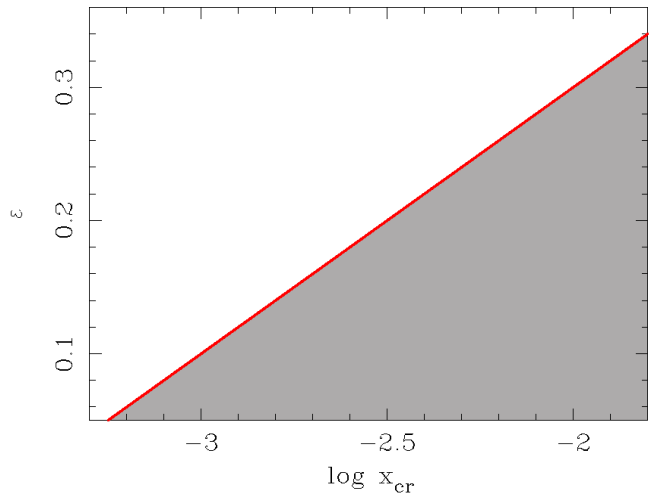,angle=270,width=0.48\textwidth}
\caption{Shaded area is the region of the parameter space ($\epsilon$,
  $x_{\rm cr}$) which must be exclude, as the total black hole mass
  density at $z=0$ is smaller than the total calculated mass density accreted
  over cosmic time.}
\label{fig:explo}
\end{figure}

\begin{figure}
\psfig{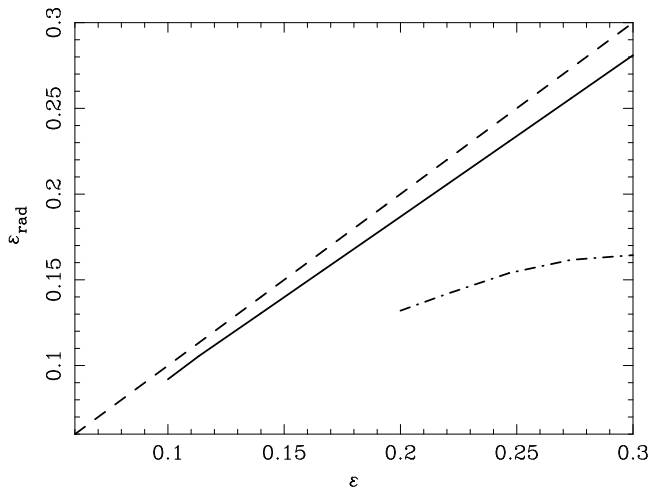}
\caption{The average radiative efficiency, $\epsilon_{\rm rad}$ 
for the evolving SMBH  
population between redshift 3.5 and 0, as a function of the accretion
efficiency $\epsilon$. Shown are the calculations done for three
values of the critical rate: $\log (x_{\rm cr})=-3.5$ (dashed line);
$-3$ (fiducial case, solid line) and $-2.5$ (dot-dashed line). Only values
of $\epsilon$ that are above the line $\epsilon\simeq 0.2 \log(x_{\rm
  cr})+0.7$ (see figure~\ref{fig:explo}) are shown.}
\label{fig:eff_radeff}
\end{figure}

A similar line of reasoning can of course be applied to our
calculations. Following the evolution of the BHMF backwards in time, we
can search for the values of the free parameters $\epsilon$ and
$x_{\rm cr}$ for which the SMBH mass density becomes negative in a
finite time. For the specific $x(\dot m)$ and bolometric
correction adopted
(see \S~\ref{sec:modes}), I found the acceptable region is bounded below by the
following empirical relation: $\epsilon\simeq 0.2 \log(x_{\rm cr})+0.7$, which
is shown in Fig.~\ref{fig:explo}, with the shaded area representing
the excluded region of the parameter space. The mean {\it radiative}
efficiency of SMBH that these calculations yield is itself a function
of $\epsilon$ and $x_{\rm cr}$, as it is shown in
figure~\ref{fig:eff_radeff}. Obviously, the higher the critical X-ray to
Eddington rate where the transition occurs, the lower the average
radiative efficiency is for any given value of $\epsilon$, as a larger
number of objects of any given $L_{\rm X}$ will be in the radiatively
inefficient regime of accretion at any time.

Figures~\ref{fig:explo} and \ref{fig:eff_radeff} seem to suggest that
 a relatively high average accretion efficiency is to be preferred,
implying a non zero average spin for the SMBH population, unless
$x_{\rm cr}$ is lower than $10^{-3}$. It should be stressed, however,
that such a constraint also depends crucially on the value of the
local black hole mass density\footnote{If $\rho_{\rm BH,0}$ is
  higher, as discussed, for example,  in
M04, then the radiative efficiency of accretion is
allowed to be substantially lower without violating any constraint},
and on the adopted bolometric
correction, and that the uncertainties in the HXLF and, to a larger
extent, in the RLF adopted here should affect the exact determination of
the accepted region of the parameter space.

In any case, it is worth stressing that the qualitative behaviour of the BHMF evolution, being
driven essentially by the evolution of the shape of the two luminosity
functions, is not modified by changes of $\epsilon$ and/or $x_{\rm
  cr}$. In particular, the anti-hierarchical character of the solution
found is a robust result of the approach presented here.

\section{Discussion}
\label{sec:disc}

In the previous section I have shown and discussed what is possible to
deduce about the evolution of the supermassive black hole mass
function  from the coupled evolution of HXLF and RLF. The limits of
this approach are a consequence of the extreme difficulty of obtaining
high redshift luminosity functions in the radio and X-ray band. Little
can be said then, about the sources evolution at higher redshifts,
where there are hints of a decline of AGN activity with redshift.
 
Theoretically, the high $z$ evolution should be mainly driven by
the gravitational growth of structures, with the 
AGN activity triggered by mergers, according to the standard
hierarchical picture (Efstathiou \& Rees 1988; Cavaliere \& Vittorini
2000; Wyithe \& Loeb 2003; Di Matteo et al. 2003; Menci et al. 2004). 
In those phases of early structure formation and evolution,
there should be plenty of gas available for
accretion due to the frequent galaxy merging events which can
effectively destabilize it and make it reach the central SMBH sphere
of influence. Black hole growth is essentially limited by the
Eddington limit (the {\it feast}, according to Small \& Blandford
1992, or the {\it self-limited regime}, Cavaliere \& Vittorini
2000). 
The shape of the black hole mass function inferred at
$z\sim 3$ (see Fig.~\ref{fig:mf}), suggesting that higher mass
black holes were already in place at that time, requires a very rapid
growth and very high accretion rates in the high density peaks of the 
cosmic density distribution 
at early times, also consistent with the trend of $\dot m$
shown in Figure~\ref{fig:zevmass}.

On the other hand, the rapid, anti-hierarchical, evolution 
of the SMBH population between
$z=0$ and $z\sim 2.5$ has always been difficult to incorporate
into standard CDM models for structure formation. 
It is interesting, in this context, to compare the main qualitative,
anti-hierarchical, picture emerging for growing black holes
with the evolution of the galaxy population. As for SMBH, the most
striking and robust evolutionary trend observed for the galaxy
population is the rapid increase of the global star formation rate
(SFR) between $z=0$ and $z\sim 1-2$ (Lilly et al. 1996; Madau et al. 1996).
Indeed, Di Matteo et al. (2003) and M04 have already shown that BHAR
and SFR histories have broadly similar shapes, with an approximately
constant ratio of about few times 10$^{-3}$, and this has also been
recently measured directly by the SDSS (Heckman et al. 2004). 
 Cowie at al. (1996) have
shown that such an evolution is caused by the smooth decline with
redshift (from $z\sim 1$ to the present) of the rest-frame K band
luminosity (an indicator of the total stellar mass) 
of galaxies that undergo rapid star formation. This
phenomenon, called `down-sizing' is the exact analogous of the
anti-hierarchical growth of supermassive black holes described here,
and has been recently confirmed by Kauffmann et al. (2004), by studying
the environmental dependence of the relation between star formation
and stellar mass in a large number of SDSS galaxies. 

What is the physical origin of such a common behavior? The combined effects
of the decrease in the galaxy interaction rate in the era of groups
and clusters formation (see e.g. Cavaliere \& Vittorini 2000),
the expansion of the universe affecting the gas cooling efficiency 
(Hernquist \& Springel 2003; Di Matteo et al. 2003), 
the progressive depletion of the cold gas
reservoirs within galaxies needed to power accretion and strong
feedback from both stars and AGN (see e.g. Wyithe \& Loeb 2003;
Granato et al. 2004), should all contribute to the observed evolution
at redshift below 3, which can be defined as that of {\it supply-limited}
accretion (Cavaliere \& Vittorini 2000), or the {\it famine} after the
feast (Small \& Blandford 1992).
Clearly, the aim of this paper is not to explore all the above theoretical
issues and predictions, rather to provide a detailed picture to
compare those predictions with. Therefore, the rest of the discussion
will be devoted to two specific results on the lifetimes and duty
cycles of active black holes and on the possible implication of the
derived SMBH growth history for the AGN feedback evolution.

\subsection{Lifetimes and duty cycles of active black holes}
In all models that try to derive the properties of the SMBH population
from the observed QSO evolution, a key element is represented by the
typical quasar lifetime or by the almost equivalent activity duty
cycle 
(see Martini 2003 and references therein).
However, the significance of these parameters is limited to the
standard case in which, on the basis of an observed luminosity
function in a specific waveband, one tries to derive the distribution
of either BH masses or accretion rates. Usually, a constant Eddington
ratio is assumed in this case, which implies that QSO are considered
as on-off switches. Then, the duty cycle is simply the fraction of
black holes active at any time, and the lifetime is the integral of
the duty cycle over the age of the universe. 

The picture discussed here is different, in that a broad distribution
of Eddington rates is not only allowed, but actually calculated for
the SMBH population at every redshift. When this is the case, a more
meaningful definition of activity lifetime is needed. I follow Steed 
\& Weinberg (2004) formulation, by first defining
the mean Eddington rate for object of mass $M_0$ at redshift $z=0$
$\langle \dot m (M_0,z) \rangle$
and then introducing the {\it mean accretion weighted lifetime} of
a SMBH with a given mass {\it today}: 
\begin{equation}
\label{eq:tau}
\tau(M_0,z) = \int_{\infty}^{z} \langle \dot m (M_0,z') \rangle 
\frac{dt}{dz'} dz',
\end{equation}

The ratio of $\tau(M_0,z)$ to the Salpeter time, $t_{\rm S}=\epsilon 
M c^2/L_{\rm Edd}= (\epsilon/0.1) 4.5 \times 10^7$ yrs, 
gives the mean number of $e$-folds of mass growth for
objects with mass $M_0$ up to redshift $z$. The ratio of $\tau(M_0,z)$ to the
Hubble time $t_{\rm Hubble}(z)=H(z)^{-1}$, instead, is a measure of the
activity duty cycle of SMBH. 

As I follow here a phenomenological approach based on observed
luminosity functions, and no information is therefore available on the
formation and early growth of the first black holes, it is interesting
here to calculate ``partial'' lifetimes in a given redshift
interval $\Delta z =(z_i,z_f)$:
\begin{equation}
\Delta \tau (M_0,\Delta z) = \int_{z_i}^{z_f} \langle \dot m (M_0,z')
\rangle
\frac{dt}{dz'} dz';
\end{equation}

\begin{figure}
\psfig{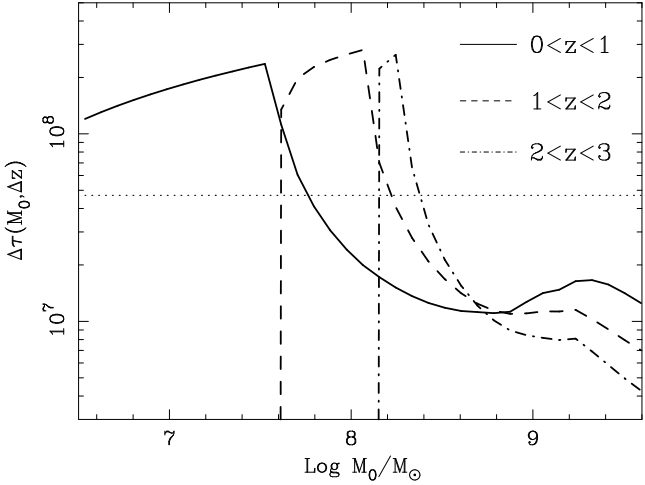}
\caption{Partial mean accretion weighted lifetimes of SMBH with mass
  today $M_0$, calculated for three different redshift intervals:
  $0<z<1$ (solid line), $1<z<2$ (dashed line) and $2<z<3$ (dot-dashed
  line). The horizontal dotted line is the Salpeter time for accretion
  efficiency of 10\%. The
accretion weighted lifetime for BH of any given mass between $0<z<3$
is the sum of the three.}
\label{fig:tau}
\end{figure}

In Figure~\ref{fig:tau}, I show $\Delta \tau (M_0,\Delta z)$ for three
redshift intervals: $0<z\le 1$; $1<z\le 2$ and $2<z\le 3$. The
accretion weighted lifetime for BH of any given mass between $0<z<3$
is of course just the sum of the three. The
anti-hierarchical nature of mass build-up in actively accreting AGN
and QSOs is again clearly illustrated by this plot. In fact, the major
growth episode of a SMBH must coincide with the period when $\Delta
\tau > t_{\rm S}$. This happens at $z<1$ for $M_0\la 10^{7.6}$, between
redshift 1 and 2 for $10^{7.6} \la M_0 \la 10^{8.2}$, and at $2<z<3$ for $
10^{8.2} \la M_0 \la 10^{8.4}$. 
Supermassive black holes with masses larger than $M_0 \sim 10^{8.5}$ today,
must have experienced their major episodes of growth at redshift higher than 3.
Black hole of lower mass today, which are also accreting at the higher
rates in the local universe (see section~\ref{sec:census}) drop below
an hypothetic seed mass (here fixed at $10^4 M_{\odot}$, but these
results do not depend strongly on this value) and effectively
``disappear'' at higher redshift. This reflects the obvious impossibility of
working out the initial condition of black hole growth 
from the local population evolved backwards once an object has
exponentiated its mass just a few times.

It also interesting to note that the objects that dominate the SMBH
mass function today, i.e. those in the range of masses around
$10^{7.5}M_{\odot}$, where $M_0
\phi_{M}(M_0,z=0)$ peaks, mainly grew around $z \sim 1$, which
is when most of the X-ray background light we see today was emitted
(Hasinger 2003).

\begin{figure}
\psfig{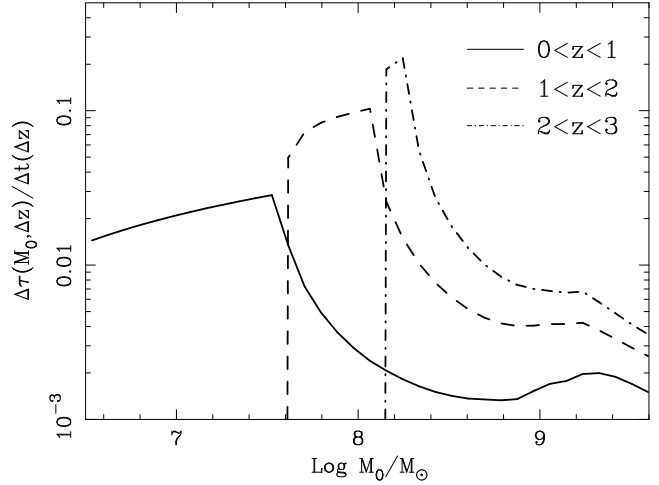}
\caption{Partial mean accretion weighted duty-cycles of SMBH with mass
  today $M_0$, calculated for three different redshift intervals:
  $0<z<1$ (solid line), $1<z<2$ (dashed line) and $2<z<3$ (dot-dashed
  line).}
\label{fig:tau_duty}
\end{figure}

The ratio $\Delta \tau (M_0,\Delta z)/\Delta t (\Delta z)$, where
$\Delta t (\Delta z)$ is the time elapsed in the redshift interval
$\Delta z$, is an indication of the ``partial'' duty cycle of a black
hole of mass $M_0$ today in that particular epoch.
This is shown in figure~\ref{fig:tau_duty} for the same redshift
intervals of fig.~\ref{fig:tau}.

\subsection{On the kinetic energy output and the history of AGN feedback}
\label{sec:feedb}
The main novelty of the constraints on black hole growth history
presented here is the opportunity to determine both mass and accretion
rate for each and every source, thanks to the fundamental plane
relationship. As I discussed in
section~\ref{sec:modes}, different modes of accretion
 are expected at different $\dot m$. In fact, not just the radiative
 output of an accreting black hole scales differently with $\dot m$
for different
 accretion modes, but also the total kinetic energy carried by the 
jets/outflows that are responsible for radio emission. 
It is possible, therefore, to derive a parallel history of the (mostly
unseen) mechanical power output from supermassive black
holes growth. 
 
To derive a scaling of the jet
kinetic power, $W_{\rm jet}$, with mass and accretion rate
for sources above and below $x_{\rm cr}$ I will proceed in the
following way. I will fist assume that the jet kinetic power at injection
is carried by internal energy, and I assume equipartition between the
jet magnetic field and the total pressure in the disc (see Heinz \&
Sunyaev 2003). Then $W_{\rm jet}\propto P_{\rm rel} R_{\rm S}^2
\propto B^2 M^2$, where  $P_{\rm rel}$ represents the pressure in
relativistic
particles at the base of the jet and  $R_{\rm S}=2GM/c^2$ is the
Schwarzschild radius. The scaling of the magnetic field, instead, can
be directly inferred from the inclination of the fundamental plane.
Equation (11) of MHD03 relates the observed
correlation coefficients of the fundamental plane relation 
to the slope of the electron distribution
in the jet, $p$, the observable
radio spectral index $\alpha_{\rm R}$, the logarithmic derivatives
of the magnetic field intensity with respect to mass and accretion
rate and the index $q$ of the $L_{\rm X}-\dot m$ relation 
(see section~\ref{sec:modes}). From that equation 
we have, assuming $p=2$ and
$\alpha_{\rm R}=0$:
\begin{eqnarray}
\frac{\partial \ln B}{\partial \dot m}&=&0.35\, q\, \xi_{\rm RX} \simeq
0.21\, q\nonumber \\
\frac{\partial \ln B}{\partial M}&=& 0.35 (\xi_{\rm RX}+\xi_{\rm
  RM})-1 \simeq -0.51,
\end{eqnarray}
where we have used the results of MHD03 $\xi_{\rm
  RX}=0.6$ and $\xi_{\rm RM}=0.78$.

Thus, we obtain the expected result that, for radiatively inefficient
flows ($q\simeq 2.3$) the total jet kinetic power is proportional 
to the physical
accretion rate only $W_{\rm jet} \propto \dot m M \propto \dot M$ 
(see Falcke \& Biermann 1996; Heinz \& Sunyaev 2003; 
Fender, Gallo \& Jonker 2003).
Whether such an output dominates the energy
budget (according to the so-called ADIOS picture, Blandford \&
Begelman 1999, and as proposed also by Fender, Gallo \& Jonker 2003) 
or not depends however on the dynamics of the innermost
disc-jet coupling and, from the observationally point of view, on the 
radiative efficiency of the jets.
On the other hand, SMBH accreting above the critical rate $x_{\rm
  cr}$, for which we have assumed $q=0.76$, should obey the scaling 
$W_{\rm jet} \propto M \dot m^{0.33}$. 

\begin{figure}
\psfig{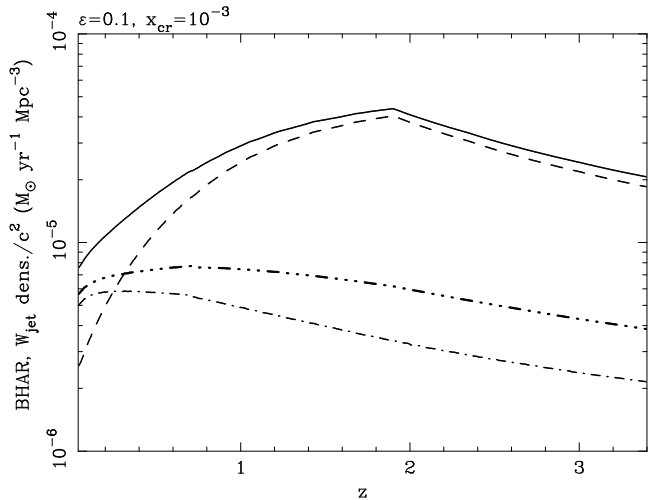}
\caption{Redshift evolution of the 
total integrated mechanical power $W_{\rm jet}/c^2$ from accreting black
holes per unit co-moving Mpc (triple dotted-dashed line). For
comparison, the evolution of the total black hole accretion rate
density (solid line) is shown together with the two separate
contributions from the sources accreting above(dashed line) and below
(dot-dashed line) the critical rate, all taken from Fig.~\ref{fig:zevmdot}.}
\label{fig:lkin}
\end{figure}

Given the above scaling relations for $W_{\rm jet}$, 
it is possible to calculate the
mechanical power output from each accreting black hole and the total
integrated one, as a function of redshift. 
For the sake of simplicity I assume that indeed the total kinetic
power of the jet/outflow from radiatively inefficient black holes
dominates as a sink of energy. Therefore, the calculated $W_{\rm jet}$
should be considered an absolute upper limit to the jet/outflow
kinetic power.  The results are shown in
Figure~\ref{fig:lkin}.

Because the relative contribution of the mechanical energy output 
is much larger in low accretion rate sources, the history of $W_{\rm
  jet}$ does not follow that of the most luminous sources, but instead
exhibit a much weaker evolution, both at low and high redshifts. Also,
because the number of low accretion rate sources increase with time
(due to the overall decrease of the average accretion rate, see
previous section), the peak of the mechanical power output 
lies at a much lower redshift than the peak of the radiative energy output
history of AGN, which should have interesting implications for the
formation and dynamical evolution of clusters of galaxies.

\section{Conclusions}
\label{sec:conc}

I have presented a new method to study the growth of accreting
supermassive black holes, based on the simultaneous evolution of the
AGN radio and hard (2-10 keV) X-ray luminosity functions.
The method is based on the locally observed trivariate correlation
between black hole mass, X-ray and radio luminosity (the so-called
fundamental plane of black hole activity, MHD03). 
Thanks to this correlation, it is possible for the first time 
to break the degeneracy between luminosity, mass and accretion rate
that affected all previous attempts to study the evolution of the supermassive
black holes population by looking at the evolution of a single AGN
luminosity function.

Here, the redshift evolution of the SMBH mass function between $z=0$ and
$z\sim 3$ is evaluated integrating backwards in time a continuity
equation for the black hole population in which the role of mergers is
neglected. The local black hole mass function, estimated by applying
the correlation between black hole mass and velocity dispersion to the
velocity distribution function of local galaxies, 
is used as a boundary condition
for the continuity equation. 
The solution to this equation is uniquely determined once
the accretion efficiency is specified, together with the function 
$L_{\rm X}/L_{\rm Edd}(\dot m)$, that links the observed Eddington
scaled hard X-ray luminosity of an accreting black hole to its
 accretion rate. For the latter, I have assumed a very general form of a double
power-law, corresponding to a radiatively inefficient regime at low
accretion rates, and a radiatively efficient one at high $\dot m$.

By comparing the local observed BHMF with the total mass accreted onto
SMBH over their history, it is possible to put simultaneous constraints on 
the accretion efficiency and on the critical value of the accretion
rate, $x_{\rm cr}$, at which the transition takes 
place between the two accretion modes.

The main results of this work are the following.
For fiducial values of the parameters ($\epsilon=0.1$ and 
$x_{\rm cr}=10^{-3}$), half ($\sim 85$\%) of the local 
black hole mass density was accumulated at redshift $z<1$ ($z<3$),
mostly in radiatively efficient episodes of accretion.
Qualitatively (i.e. independently on the values of these two
parameters), the evolution of the black hole mass
function between $z=0$ and $z\sim 3$ shows clear signs of an 
{\it anti-hierarchical} behaviour. This is a purely phenomenological
assessment, and reflects the fact that, while the majority of the 
most massive objects
($M\ga 10^9$) were already in place at $z\sim 3$, lower mass ones
mainly grew at progressively lower redshift, so that 
the average black hole mass increases with
increasing redshift. On the other hand, the average accretion rate decreases
towards lower redshift. Therefore, sources in the RIAF regime
of accretion only begin to dominate the comoving accretion energy
density in the universe at $z<1$ (with the exact value of $z$ 
depending on $x_{\rm cr}$), while at the peak of the black hole
accretion rate history, radiatively efficient accretion dominates by
almost an order of magnitude. By carefully evaluating the contributions
to the total black hole accretion energy density from the different
modes of accretion as a function of redshift, I show how to derive a
physically motivated AGN feedback history. This, as well as the
anti-hierarchical behaviour of SMBH growth described above, 
may be of some importance for cosmological models of structure 
formation in the universe.

\section*{Acknowledgments}
I thank Xuelei Chen, Tiziana Di Matteo, Sebastian Heinz and Susumu Inoue for
helpful discussions, and the anonymous referee for his/her useful comments.

\appendix
\section{Analytical approximation of the X-ray luminosity function}
I adopt here the functional form for the HXLF described in Ueda et
al. (2003), 
which has the following analytic approximation:
\begin{equation}
\phi_{\rm X}(L_{\rm X},z)=\phi_{\rm X}(L_{\rm x},0)\eta(z,L_{\rm X}),
\end{equation}
where the local X-ray luminosity function is expressed as smoothly
connected double power-law:
\begin{equation}
\phi_{\rm X}(L_{\rm X},z=0)=A[(L_{\rm X}/L_*)^{\gamma_1}+(L_{\rm X}/L_*)^{\gamma_2}]^{-1},
\end{equation}
while the evolutionary part is expressed as
\begin{equation}
\eta(z,L_{\rm x}) = \left\{
        \begin{array}{ll}
        (1+z)^{p_1} & \hbox{$ z < z_{\rm c}(L_{\rm X})$} \\
        \eta(z_{\rm c})[(1+z)/(1+z_{\rm c}(L_{\rm X}))]^{p_2}   &
        \hbox{ $z \ge z_{\rm c}(L_{\rm X})$}   \\
        \end{array}\right.\;
\end{equation}
and
\begin{equation}
z_{\rm c}(L_{\rm x}) = \left\{
        \begin{array}{ll}
        z_{\rm c}^* & \hbox{$ L_{\rm X} \ge L_a$} \\
        z_{\rm c}^*(L_{\rm X}/L_a)^{\alpha} &
        \hbox{ $L_{\rm X}< L_a$}   \\
        \end{array}\right.\;
\end{equation}
For a $\Lambda$CDM universe, the best fit parameters are: $A=5.04\pm
0.33 \times 10^{-6} h_{70}^3$ Mpc$^{-3}$;
$L_*=10^{\left(43.94^{+0.21}_{-0.26}\right)} h_{70}^{-2}$ erg s$^{-1}$;
$\gamma_1=0.86 \pm 0.15$; $\gamma_2= 2.23 \pm 0.13$; $p_1=4.23 \pm
0.39$; $p_2=-1.5$ (fixed); $z_{\rm c}^*=1.9$ (fixed); 
$L_a=10^{44.6} h_{70}^{-2}$ erg s$^{-1}$
(fixed); $\alpha=0.335\pm 0.070$.
 
\section{Analytical approximation of the radio luminosity function}
The 5GHz RLF is derived from the low-frequency (151 MHz) one of
Willott et al. (2001), assuming a fixed radio spectral index of
$\alpha_{\rm R}=0.7$ to rescale the luminosities.

The analytical approximation is that of model C of Willott et
al.(2001), and is given by a sum of two differently evolving
populations (a low and a high luminosity one): 
\begin{equation}
\phi_{{\rm R},151}(L_{151},z)=\phi_{\rm R,l}+\phi_{\rm R,h}.
\end{equation}

These two populations evolve differently with redshift, according to
the following expressions:
\begin{equation}
\phi_{\rm R,l}=\left\{ 
  \begin{array}{ll}
    \phi_{\rm R,l}^0 \left(\frac{L_{151}}{L_{\rm
      l,*}}\right)^{-\alpha_{\rm l}} \exp \left(\frac{-L_{151}}{L_{\rm
      l,*}}\right) (1+z)^{k_{\rm l}}; & z<z_{{\rm l}0} \\
    \phi_{\rm R,l}^0 \left(\frac{L_{151}}{L_{\rm
      l,*}}\right)^{-\alpha_{\rm l}} \exp \left(\frac{-L_{151}}{L_{\rm
      l,*}}\right) (1+z_{{\rm l}0})^{k_{\rm l}}; &  z \ge z_{{\rm l}0} \\
\end{array} \right.
\end{equation}
\begin{equation}
\phi_{\rm R,h}=\phi_{\rm R,h}^0 \left(\frac{L_{151}}{L_{\rm
      h,*}}\right)^{-\alpha_{\rm h}} \exp \left(\frac{-L_{\rm
      h,*}}{L_{151}}\right) f_{\rm h}(z).
\end{equation}
The high luminosity redshift evolution $f_{\rm h}(z)$ is given by:
\begin{equation}
f_{\rm h}(z)= \left\{
  \begin{array}{ll}
    \exp \left[ -\frac{1}{2} \left( \frac{z-z_{{\rm h}0}}{z_{\rm
    h1}}\right)^2\right] & z<z_{{\rm h}0} \\
    \exp \left[ -\frac{1}{2} \left( \frac{z-z_{{\rm h}0}}{z_{\rm
    h2}}\right)^2\right] & z \ge z_{{\rm h}0} \\
\end{array} \right.
\end{equation}

The best-fitting parameters to the observed luminosity function are:
$\log \phi_{\rm R,l}^0 = -7.12 ^{+0.10}_{-0.11}$, $\alpha_{\rm
  l}=0.539 \pm 0.02$, $\log L_{\rm l,*} = 26.10^{+0.08}_{-0.09}$, 
$z_{{\rm l}0}=0.71 \pm 0.10$, $k_{\rm l}=4.30^{+0.57}_{-0.55}$, 
$ \log \phi_{\rm R,h}^0 = -6.20 ^{+0.09}_{-0.11}$, $\alpha_{\rm
  h}=2.27^{+0.12}_{-0.11}$, $\log L_{\rm h,*}=26.95^{+0.11}_{-0.10}$,
$z_{{\rm h}0}= 1.91 \pm 0.16$, $z_{\rm h1}= 0.56 \pm 0.05$, $z_{\rm
  h2}= 1.38 ^{+0.52}_{0.28}$.
\bsp

\label{lastpage}

\end{document}